\documentstyle[preprint,aps,floats,tighten]{revtex}
\input{psfig.tex}
\begin{document}
\draft
\preprint{\vbox{\hbox{CU-TP-736} 
                \hbox{CAL-593}
                \hbox{astro-ph/9601147}
}}

\title{A Low-Density Closed Universe}

\author{Marc Kamionkowski\footnote{kamion@phys.columbia.edu}\
and Nicolaos Toumbas}
\address{Department of Physics, Columbia University,
New York, New York~~10027}
\date{January 1996}
\maketitle

\begin{abstract}
Matter with an equation of state $p=-\rho/3$ may arise in
certain scalar field theories, and the energy density of this
matter decreases as $a^{-2}$ with the scale factor $a$ of
the Universe.  In this case, the Universe could be closed but
still have a nonrelativistic-matter density $\Omega_0<1$.
Furthermore, the cosmic microwave background could come from a
causally-connected region at the other side of the
Universe.  This model is currently viable and might be tested by
a host of forthcoming observations.
\end{abstract}

Of the three possibilities, a closed Universe receives far
less attention in the current literature than an open or a flat
Universe. Observations that find a matter density less than
critical suggest an open Universe.  Theoretical arguments,
such as the Dicke coincidence and inflation, favor a flat Universe.
However, there are heuristic arguments for a closed
Universe that involve, for example, consistency of quantum
field theories on a compact space or the idea that it is easier
to create a finite Universe with zero energy, charge, and angular momentum.
Even so, given the observations, it requires some {\it chutzpah}
to suggest that the matter density is greater than critical.
For these reasons, models that are closed by virtue of a
cosmological constant ($\Lambda$) have been recently considered
\cite{scottwhite}.  In this paper, we consider a variation: a
low-density closed Universe, which at low redshifts
is entirely indistinguishable from a standard
open Friedmann-Robertson-Walker (FRW) Universe with
the same non-relativistic matter density.

If some form of matter with an equation of state $p=-\rho/3$
exists, then its energy density decreases with the
scale factor $a$ of the Universe as $a^{-2}$ and thus
mimics a negative-curvature term in the Friedmann equation.
In this case, the Universe could be closed and still have
a nonrelativistic-matter density $\Omega_0<1$.

In fact, the energy density contributed by a scalar field with a
uniform gradient-energy density would scale as $a^{-2}$.
However, such a
scalar-field configuration would collapse within a Hubble
time unless it was somehow stabilized.  Davis \cite{davisone}
pointed out that
if there was a manifold of degenerate vacua with nontrivial
mappings into the three-sphere [which could be accomplished if
there was a global symmetry $G$ broken to a subgroup $H$ with
$\pi_3(G/H) \neq 1$], then a texture---a topological defect with
uniform gradient-energy density---would be stabilized provided
that it was wound around a closed Universe
\cite{davisone}.  Non-intersecting strings would also provide an
energy density that scales as $a^{-2}$.

Moreover, if the energy density contributed
by the texture is chosen properly, the observed cosmic microwave
background (CMB) comes from a causally-connected patch at
the antipode of the closed Universe
\cite{daviscmb}.\footnote{This could similarly be accomplished
with $\Lambda\neq0$, but these models are likely ruled out by
lensing statistics \cite{scottwhite}.}
Although the homogeneity problem is still not addressed, 
we find it
illustrative and interesting that one can still construct a
viable model, which looks remarkably like an open Universe at low
redshifts, even though the largest-scale structure differs
dramatically. 

The Friedmann equation for a closed Universe
with nonrelativistic matter and some other
form of matter (perhaps a stable texture) with an equation of
state $p=-\rho/3$ is
\begin{eqnarray}
     H^2 &\equiv& \left({\dot a \over a}\right)^2 = {8\pi G \over 3}
     \rho_m + {\gamma - 1 \over a^2} \nonumber \\
     &=&H_0^2 [ \Omega_0 (1+z)^3 +(1-\Omega_0)(1+z)^2]
     \equiv H_0^2 [E(z)]^2,
\label{friedmann}
\end{eqnarray}
where $H=\dot a/a$ is the Hubble parameter (and the dot denotes
derivative with respect to time), $z=(a_0/a)-1$ is the redshift,
$G$ is Newton's gravitational constant, $\rho_m$ is the density of
nonrelativistic matter, and $\gamma$ is a parameter that
quantifies the contribution of the energy density of the
texture.  The second line defines the function $E(z)$.
This is exactly the same as the Friedmann equation for an open
Universe with the same $\Omega_0$, so this closed Universe has
the same expansion dynamics.  At the current
epoch (denoted by the subscript ``0''),
\begin{equation}
     \Omega_0=1 + {1-\gamma \over a^2 H^2} = 1-\Omega_t +
     {1\over a_0^2 H_0^2},
\label{Omegaequation}
\end{equation}
where $\Omega_t=\gamma(a_0 H_0)^{-2}$ is the contribution of the
texture to closure density today.  So, $\Omega_0<1$ if
$\gamma>1$ even though the Universe is closed, and we require
that $\Omega_t + \Omega_0>1$.

If the metric of a closed Universe is written as
\begin{equation}
     ds^2= dt^2 - a^2(t) \left[ d\chi^2 + \sin^2 \chi (
     d\theta^2 + \sin^2 \theta d\phi^2) \right],
\end{equation}
then the polar-coordinate distance between a source at a
redshift $z_1$ and another source along the same line of sight
at a redshift $z_2$ (for $\Omega_0<1$) is
\begin{equation}
     \chi_2 -\chi_1 = \sqrt{\Omega_0 + \Omega_t -1}
     \int_{z_1}^{z_2} \, {dz \over E(z)}.
\label{chiequation}
\end{equation}
If $\Omega_t$ is chosen such that the
polar-coordinate distance of the CMB surface of last
scatter is $\chi_{LS}\simeq \pi$, then the CMB we observe comes
from a causally-connected patch at the antipode of the
Universe.  Since this Universe expands forever, we
could also choose $\chi_{LS}\simeq2\pi$, in which case the CMB
photons have traveled precisely once around the Universe.
This introduces the intriguing possibility
that when we observe the CMB we are looking at the {\it local}
(rather than some distant) region of the Universe as it was at a
redshift $z\simeq1100$.  In fact, for
$\chi_{LS}\simeq n\pi$ with $n=1,2,3,...$, CMB photons have
traveled $n/2$ times around the Universe, and the CMB comes from a
causally-connected patch on the other side of the Universe
(for $n$ odd) or from the local neighborhood (for $n$ even).
From Eq.\ \ref{chiequation}, the condition on $\Omega_t$ for
$\chi_{LS}=n\pi$ is
\begin{equation}
     \Omega_t=\left[ {n\pi \sqrt{1-\Omega_0} \over {\rm arcsinh}
     ( 2 \sqrt{1-\Omega_0}/\Omega_0)} \right]^2 +1 - \Omega_0.
\label{conditioneqn}
\end{equation}
For $n=1$ ($n=2$), $\Omega_t$ increases from 1.6 to 2.5 (4 to
10) for $\Omega_0$ between 0.1 and 1.

Is this a realistic possibility?  For $n\geq2$, it requires a
radius of curvature for the Universe that is probably
too small to be 
consistent with observations.  The $n=1$ case is still
consistent with our current knowledge of the Universe.  However,
forthcoming
observations may be used to distinguish it from a standard open
Universe, as we now explain.\footnote{For an excellent review of
many classical cosmological tests, see Ref. \cite{peebles}.}

Since the expansion dynamics is the same as for an open FRW
Universe, quantities that depend only on the
expansion, such as the deceleration parameter, the age of the
Universe, or the distribution of quasar absorption-line redshifts,
do not probe $\Omega_t$.  Furthermore, the growth of
density perturbations is the same as in a standard open Universe, so
dynamical measurements of $\Omega_0$ (e.g., from
peculiar-velocity flows) will also be insensitive
to $\Omega_t$.  Effects due to geometry arise only at ${\cal
O}(z^3)$ since $\sin\chi$ and $\sinh\chi$ differ only at ${\cal
O}(\chi^3)$; therefore, this Universe will differ from an
open Universe only at $z\gtrsim1$.

Ergo, we now turn to cosmological tests that probe the
geometry of the Universe.  Underlying these is the
angular-diameter distance between a source at a redshift $z_2$
and a redshift $z_1<z_2$,
\begin{equation}
     d_A(z_1,z_2)= { \sin(\chi_2-\chi_1) \over (1+z_2) H_0
     \sqrt{\Omega_0+\Omega_t-1}}.
\label{angulardiametereqn}
\end{equation}
The angular size of an object of proper length $l$ at
a redshift $z$ is $\theta\simeq l/d_A(0,z)$.  Consider first the
case where $\Omega_t$ is fixed by $n=2$.  Then the antipode
$\chi=\pi$ of the Universe must be at some redshift $z_a<1100$.
One finds that  $z_a\lesssim5$ for $\Omega_0\gtrsim0.3$, and
therefore, the angular sizes of the highest-redshift quasars
must be very large.  Additional arguments against an
antipode at $z\lesssim5$ for a closed $\Lambda$ Universe have
been given Refs. \cite{gott}.  These arguments probably apply
to the model considered here, although we have not done a
complete analysis.  Therefore, a closed Universe with $n\geq2$
is highly unlikely and we pursue it no further.

\begin{figure}[htbp]
\centerline{\psfig{file=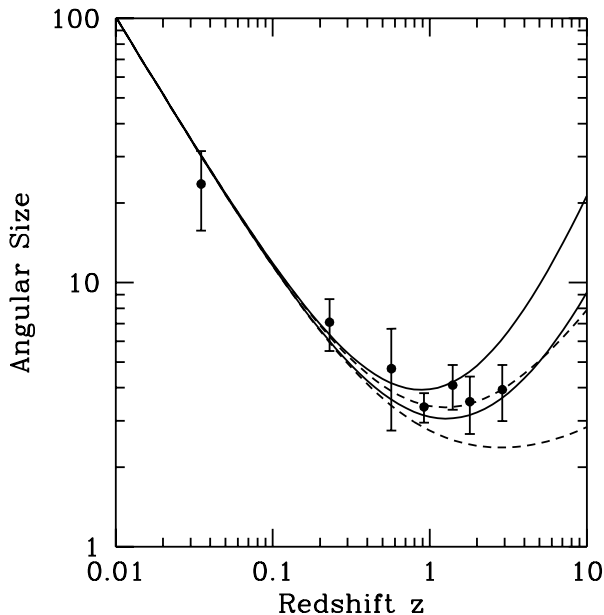,width=3.3in}}
\bigskip
\caption{
  The angular size of an object of proper length $l$ (in units
  of $l H_0$) for the closed Universe (solid curves) and for an
  open and flat FRW Universe (dashed curves).  In each case, the
  upper curves are for $\Omega_0=1$ and the lower curves are for
  $\Omega_0=0.1$.  The points are from Ref. [6].
}
\label{angleplot}
\end{figure}

In Fig.\ \ref{angleplot}, we plot the angular size as a function
of redshift fixing $\Omega_t$ so that the CMB comes from the
antipode [i.e., Eq.\ \ref{conditioneqn} with $n=1$].  We
also plot the results for a FRW Universe.  The Figure shows that
the angular sizes in a flat matter-dominated Universe can be
roughly similar to those in a low-density closed Universe.  Therefore,
an analysis of the angular sizes of some compact radio sources,
which shows consistency with a flat Universe \cite{radios},
may also be consistent with a low-density closed Universe.
Proper-motion distances of superluminal jets in radio sources at
large redshift may provide essentially the same probe as do
flux--redshift relations.  The common caveat
is that evolutionary effects must be understood if these are to
provide reliable cosmological tests.  It has been proposed that
these effects may conceivably be understood well enough to
discriminate between open and flat $\Lambda$ models
\cite{kraussschramm}.  Fig.\ \ref{angleplot} illustrates,
however, that the difference between the angular sizes for the
FRW Universe and the closed model for the same value of
$\Omega_0$ is quite a bit more dramatic than the difference
between open FRW and flat $\Lambda$ models (c.f., Fig.\ 13.5 in
Ref. \cite{peebles}).  Therefore, if the angle-redshift relation
can distinguish open and flat $\Lambda$ models, then the
distinction between these and the closed model will be even
clearer.

\begin{figure}[htbp]
\centerline{\psfig{file=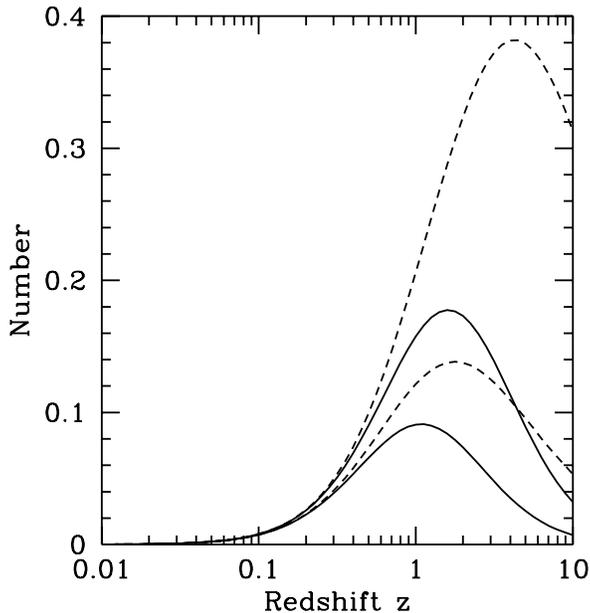,width=3.3in}}
\bigskip
\caption{
  The differential number of galaxies per unit redshift per
  steradian in
  units of $n_0 H_0^{-3}$ for the closed Universe (solid curves)
  and the open FRW Universe (dashed curves).  The upper
  curves are for $\Omega_0=0.3$ and the lower curves are for
  $\Omega_0=1$.
}
\label{numbersfig}
\end{figure}

Another classical cosmological test is the number-redshift
relation.  In the low-density closed Universe, the differential
number of galaxies per steradian per unit redshift is,
\begin{equation}
     {dN_{\rm gal} \over dz d\Omega} = {n_0  \sin^2[\chi(z)]
     \over H_0^3 (\Omega_0 +\Omega_t-1)E(z)},
\end{equation}
where $n_0$ is the local number density of galaxies, and the number
per comoving volume is assumed to remain constant.  In
Fig.\ \ref{numbersfig}, we plot the
number-redshift relation for the low-density closed Universe
with $\Omega_t$ chosen so that the CMB comes from the antipode
and for standard open and flat FRW models.  The Figure shows that
an application of this test, which finds values of $\Omega_0$
near unity in a FRW Universe \cite{lohspillar},
can also be consistent with a low-density closed Universe.
However, galactic evolutionary effects are realistically quite
significant, so this remains a controversial test.

\begin{figure}[htbp]
\centerline{\psfig{file=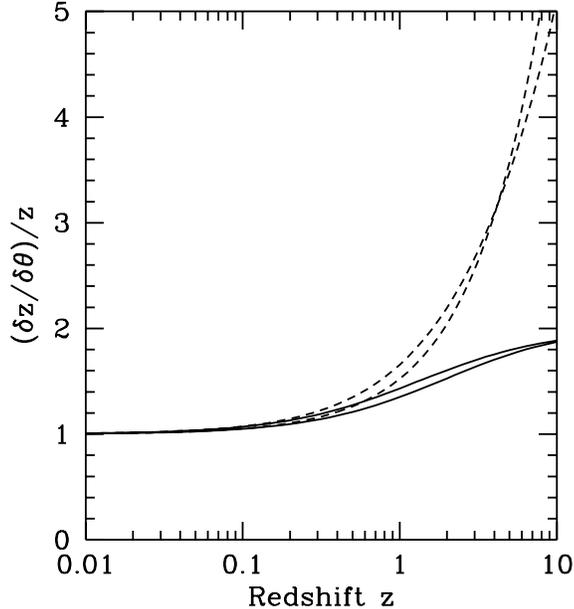,width=3.3in}}
\bigskip
\caption{
  Plot of $(\delta z/\delta\theta)/z$ for the closed Universe
  (solid curves) and the open FRW Universe (dashed curves).
  Shown are curves for both $\Omega_0=0.1$ and $\Omega_0=1$.
}
\label{dzdthetafig}
\end{figure}

A test for $\Lambda$ discussed by Alcock and
Paczy\'nski \cite{alcock} may also be an especially effective
probe of $\Omega_t$.  The redshift thickness $\delta z$ and
angular size $\delta \theta$ of a roughly spherical structure
that grows with the expansion of the Universe will have a ratio
\begin{equation}
     {1 \over z}{\delta z \over \delta \theta}=
     {E(z)\sin[\chi(z)] \over z \sqrt{\Omega_0+\Omega_t-1}}.
\end{equation}
As shown in Fig.\ \ref{dzdthetafig}, this function is
significantly lower in a low-density closed Universe than it is
in an open Universe (and in a $\Lambda$ Universe;
c.f., Fig.\ 13.9 in Ref. \cite{peebles}).  Furthermore, it depends
only very weakly on the value of $\Omega_0$ and
therefore provides an $\Omega_0$-independent determination of
the geometry.  A precise measurement may be feasible with
forthcoming quasar surveys \cite{phillips}.

We have also checked the probability for gravitational lensing
of sources at high redshift.  This test provides perhaps the
strongest constraint on $\Lambda$ models \cite{gravlenses}, and
makes it unlikely that the CMB comes from the antipode of a Universe
that is closed with the addition of a cosmological constant
\cite{scottwhite}.  The probability for lensing of a source at
redshift $z_s$ for $\Omega_0<1$ and $\Omega_t+\Omega_0>1$
relative to the fiducial case of a standard flat Universe is
\begin{eqnarray}
     P_{\rm lens} & = & {15\over4} \left[ 1- {1\over
     (1+z_s)^{1/2} } \right]^{-3} \nonumber \\
         & & \times \int_{0}^{z_s}\, {(1+z)^2 \over E(z)} \left[
	 {d_A(0,z)d_A(z,z_s) \over d_A(0,z_s)} \right]^2\,dz.
\end{eqnarray}
The current observational
constraint is roughly $P_{\rm lens}\lesssim5$.  If $\Omega_t$
is chosen so that the CMB comes from the antipode, then $P_{\rm
lens}<2.5$ for $0<\Omega_0<1$.  Hence the model is consistent with
current data and is likely to remain so.

Finally, if ours is actually a low-density closed Universe, it
will probably have a dramatic signature in the
anisotropy spectrum of the CMB, especially if the CMB
comes from the
antipode of the Universe.  Although the detailed shape of the
anisotropy spectrum depends on a specific model for
structure formation, it quite generically has structure (known
as ``Doppler peaks'') on angular scales smaller than that
subtended by the horizon at the surface of last scatter.  The
angle subtended by the horizon at last scatter depends on the
cosmological model; in a standard FRW Universe, it is
$\theta_{LS} \simeq \Omega^{1/2}\,1^\circ$.  Therefore,
measurement of the location of the first Doppler peak
provides a determination of the geometry of
the Universe \cite{kss}, and with forthcoming all-sky CMB maps
with sub-degree angular resolution, this measurement may be
quite precise \cite{jkks}.

The angular scale subtended by the horizon in a
low-density closed Universe may be approximated by
\begin{equation}
     \theta_{LS} \simeq 2^\circ\, {\sqrt{\Omega_0+\Omega_t-1}
     \over \Omega_0^{1/2} \sin\chi_{LS}},
\end{equation}
when $\theta_{LS}$ evaluates to small angles; otherwise,
$\theta_{LS}={\cal O}(\pi)$.  Here,
\begin{equation}
     \chi_{LS}=\sqrt{\Omega_0 +\Omega_t-1 \over 1-\Omega_0} {\rm
     arcsinh} \left( {2 \sqrt{1-\Omega_0} \over \Omega_0} \right)
\end{equation}
is the polar-coordinate distance traversed by the CMB photons since last
scatter.  As expected, this is always larger than $\theta_{LS}$
for a flat or open FRW Universe.  Moreover, if
$\chi_{LS}\simeq \pi$, the Doppler-peak structure of the CMB is
shifted to the largest angular scales, and the suppression of
CMB anisotropies due to Silk damping is also shifted to larger
angular scales. The precise shift depends on exactly how close the
last-scattering surface is to the antipode.\footnote{For
example, the anisotropy spectrum might resemble those shown in
Fig.\ 6 of Ref. \cite{scottwhite} for the analogous case with a
cosmological constant for a
flat scale-invariant spectrum of density perturbations.  However,
the overall tilt of the spectrum depends on the model of
primordial perturbations and could therefore be considerably
different.}
It is almost certain that these signatures will be
distinguishable in forthcoming CMB maps if they are indeed
there.

Although there is no horizon problem in this model, at
earlier or later epochs, the CMB is not generally at the
antipode.  Furthermore, the homogeneity of the Universe is not
necessarily explained even if the CMB comes from a
causally-connected region.  Even so, it is worth noting that one
can construct a viable model, which is indistinguishable from
an open Universe at redshifts $z\lesssim1$, with a closed
geometry.  Furthermore, the model will be tested by forthcoming
observations of the Universe at large redshifts, especially
through angular sizes, $\delta z/\delta\theta$, and the CMB.

We have focussed in our numerical work on the case
where $\Omega_t$ is such that the CMB comes precisely
from the antipode.  However, one could explore other values of
$\Omega_t$, perhaps within the context of flat inflationary
models.

Finally, what about the homogeneous matter with an energy density
which scales as $a^{-2}$?  If this is due to a topologically
stabilized scalar-field configuration, as discussed above, then
the symmetry-breaking scale must be of order the Planck scale if
$\Omega_t$ is of order unity.  Furthermore, the global symmetry
must be {\it exact}.  This model would therefore have significant
implications for Planck-scale physics if verified \cite{global}.

\bigskip

We thank D. Helfand and D. Spergel for useful comments.  This
work was supported in part by the D.O.E. under contract
DEFG02-92-ER 40699 and by NASA under contract NAG5-3091.

\end{document}